# Older LGBT+ and Blockchain in Healthcare: A Value Sensitive Design Perspective

**Adam Poulsen**
Brain and Mind Centre,
The University of Sydney
Sydney, NSW Australia
adam.poulsen@sydney.edu.au
orcid.org/0000-0002-0001-3894

**Eduard Fosch-Villaronga**
eLaw Center for Law and Digital
Technologies, Leiden University
Leiden, the Netherlands
e.fosch.villaronga@law.leidenuniv.nl
orcid.org/0000-0002-8325-5871

**Abstract**

Most algorithms deployed in healthcare do not consider gender and sex despite the effect they have on individuals' health differences. Missing these dimensions in healthcare information systems is a point of concern, as neglecting these aspects will inevitably perpetuate existing biases, produce far from optimal results, and may generate diagnosis errors. An often-overlooked community with distinct care values and needs are LGBT+ older adults, which has traditionally been under-surveyed in healthcare and technology design. This paper investigates the implications of missing gender and sex considerations in distributed ledger technologies for LGBT+ older adults. By using the value sensitive design methodology, our contribution shows that many value meanings dear to marginalized communities are not considered in the design of the blockchain, such as LGBT+ older adults' interpretations of trust, privacy, and security. By highlighting the LGBT+ older population values, our contribution alerts us to the potential discriminatory implications of these technologies, which do not consider the gender and sex differences of marginalized, silent populations. Focusing on one community throughout – LGBT+ older adults – we emphasize the need for a holistic, value sensitive design approach for the development of ledger technologies for healthcare, including the values of everyone within the healthcare ecosystem.

# 1. Introduction

Blockchain in healthcare, including aged care, is presented as an enabling technology used to create greater interoperability between existing information systems (IS), bringing benefits such as transparency, data accuracy, traceability, and reliability. By adopting blockchain in healthcare, some argue that it will increase trust (or reliability), security, and privacy health information (Liang, Zhao, Shetty, Liu, & Li, 2017; Agbo et al., 2019). Blockchain in care domains is a contentious topic, while much work is being done on realising blockchain-enabled health information systems (BCHIS), many others doubt the benefits. Indeed, as technology is not value-neutral (Friedman & Hendry, 2019), technologies such as distributed ledger technologies have positive and negative impacts on the values within and arising from the ecosystem where these technologies are implemented (Pazaitis, De Filippi, & Kostakis, 2017).

The advent of BCHIS has many positive applications across healthcare, including improved electronic health record management, health data analytics, and cross-institution data sharing. However, the distributed invention reduced everything to bits of data, leaving aside other fundamental aspects inherent to technological ecosystems, including human values. Indeed, the introduction of ledger technologies, artificial intelligence (AI), and robotics in healthcare are expected to increase productivity and resource efficiency, as has happened in the industrial and retail sectors (Cresswell, Cunningham-Burley, & Sheikh, 2018). The reality, however, is that these technologies have a significant impact on human values that should be taken into account beforehand, not as an afterthought (Friedman & Hendry, 2019).

The interplay between values, aged care, and distributed ledger technologies has not received much attention, unlike the primary focus has been on cryptocurrencies, smart contracts, or smart properties (De Filippi & Loveluck, 2016). The *lex cryptographica* and the legal analysis have also revolved around smart contracts, intellectual property, and data protection (Bacon et al., 2017), working on the impacts of such technologies on the law. There is increasing use of distributed ledger technologies for healthcare, and the number of publications is growing (Prokofieva & Miah, 2019). Although healthcare is a sensitive domain of application with very distinct values, a thorough analysis considering human values and value meanings is lacking.

We address this gap by investigating the impact of using and developing distributed ledger technologies (or blockchain) in healthcare for vulnerable populations. We focus on a specific community in care with its own set of value meanings: LGBT+ older adults[1]. LGBT+ older adults suffer from extreme loneliness due to historical discrimination, lack of social acceptance, isolation, and underestimated feelings (Hughes, 2016). Each cultural group has distinct value priorities, meanings, conceptualizations, and definitions (Burchum, 2002). LGBT+ older adults are no different (Waling & Roffee, 2017). Applying the value sensitive design methodology, our contribution shows that many values dear to marginalized communities are not considered by the technical blockchain literature. Moreover, by highlighting this community's under-surveyed values, our contribution shows the potential discriminatory implications of these technologies, which do not consider silent aging populations (Poulsen, Fosch-Villaronga & Søraa, 2020). By considering the values of a particular group, we emphasize the importance of recognizing the diversity of values among the aging population to avoid system bias and acknowledge the required cultural sensitivity in care (Santana et al., 2018).

After a brief introduction, in the second section, we describe what distributed ledger technologies are, which healthcare applications exist, and how systems may inadvertently violate privacy, reinforce social prejudices, and discriminate against different users. In the third section, we introduce the lens through which we address this topic, i.e., the value sensitive design (VSD) method, followed by an introduction of the older LGBT+ community case study. We finish the article by conceptualizing the values of this community concerning blockchain to stress the importance of accounting for the diversity of values in healthcare technology.

## 2. Literature Review

*2.1. The blockchain*

In 2008, Satoshi Nakamoto introduced *blockchain*. Blockchain was not a major technological innovation, but incremental development of existing technologies such as private-public key encryption and peer-to-peer networks developed in the 1970s, consensus mechanisms, and decentralized, distributed data storage developed later (Wright & De Filippi, 2015). Blockchain is an open, public ledger stored on many decentralized nodes to support for intermediary-free transactions called decentralized, trustless peer-to-peer transactions (Prokofieva & Miah, 2019). That is, participants in the network need not trust each other (Bacon et al., 2017). Iansiti (2017) defined blockchain as "an open, distributed ledger that can record transactions between two parties efficiently and in a verifiable and permanent way," while Bacon et al. (2017) referred to the blockchain as "a specific type of database that uses certain cryptographic functions to achieve the requirements of data integrity and identity authentication."

---

[1] Lesbian-Gay-Bisexual-Transgender-Intersex-Queer+others, aged 65 and above

Distributed ledger technologies have several advantages over existing systems, such as traditional distributed database management systems like Structured Query Language-based systems and NoSQL-based systems (Kuo, Kim, & Ohno-Machado, 2017) and traditional ledger-based system architectures which relied on a central or trusted third-party ledger (Hughes et al., 2019). It allows data integrity by creating persistent records of relevant transactions and identity authentication without involving intermediaries, i.e., it eliminates the middleman (Bacon et al., 2017). Before blockchain, it was impossible to organize individual activities over the Internet without a centralized body and ensure no interference. It is considered a tamper-proof system unless the controlling parties decide to alter the history of the blockchain (Atzei, Bartoletti, & Cimoli, 2017), supported by an append-only data structure and a data verification feature via consensus protocols. These protocols remove the risk of duplicate entry or fraud (Prokofieva & Miah, 2019). It also heavily relies on cryptography to secure the data ledgers with both the current and its adjacent completed block involved in the cryptography process (Prokofieva & Miah, 2019). Registered transactions cannot be altered unless the whole chain of actual recorded transactions is changed with mutual consensus, ensuring that the data stored on the blockchain is reliable and not edited without traceability. In this way, blockchain "provides a way for people to agree on a particular state of affairs and record that agreement in a secure and verifiable manner" (Wright & De Filippi, 2015).

Additionally, with blockchain, all transactions are recorded in chronological order, time-stamped, and visible to everyone on the blockchain, making it transparent. Users can remain anonymous as transactions occur between encrypted alphanumeric addresses. Users can programmatically set up algorithms to trigger transactions (Crosby, 2016). The origin of any ledger can be backtracked along the chain. The ledger is distributed across every single node in the blockchain, making it highly distributed. All transactions stored in the blocks are contained in the chain, making it decentralized and guaranteeing data recovery. These features make blockchain technology extremely useful, with salient advantages of transparency, decentralization, and security in many sectors, including government (Warkentin & Orgeron, 2020) and business (Frizzo-Barker et al., 2020). Companies have been using blockchain to track items through complex supply chains for a while (Iansiti & Lakhani, 2017), and its adoption has been critical for meeting the citizens' demands. Schuetz and Venkatesh (2019) showed, for instance, that by increasing financial inclusion in rural areas, blockchain technology has the potential to connect rural populations to supply chains. In the energy sector, distributed ledger technologies may advance energy practices and processes' efficiency, advance decentralized generation, and allow peer-to-peer energy trading (Andoni et al., 2019).

Although the recent hype of blockchain and its transition to AI, the research on blockchain adoption is still in its infancy. Until now, the adoption of such complex technology research highlighted professionals' expectations (Queiroz & Wamba, 2019), but it still lacks a greater understanding of how particular populations, such as older adults, would see the benefits of applying such a technology. Furthermore, there is a lack of literature on the challenges these technologies pose to users and with which remedies they are empowered to mitigate those risks.

*2.2. Healthcare and blockchain*

Medicine falls behind other sectors in embracing new technologies, as it is a highly complex, sensitive, and regulated sector (Hoy, 2017). Together with the continuous shortage in healthcare budgets, the difficult choice in expenditure allocation, and the potential adverse effects on users, the adoption of healthcare technologies is not straightforward. Thus, there are no significant surveys investigating distributed ledger technology applications in healthcare. However, the literature stresses that they could address urging issues in healthcare, such as fragmented records and hard access to patient health information, thanks to fundamental features like immutability, decentralization, and transparency (Zhang et al., 2018). There are numerous systems applying blockchain in healthcare found in the literature. Some examples include mobile healthcare applications with cloud storage (Liang et al., 2017); blockchain-enabled diagnosis and treatment systems (Wang et al., 2018); cloud-based health resource sharing for diagnosis (Zhu, Shi, & Lu, 2019); electronic health record management (Khatoon, 2020); drug supply chain governance (Tseng, Liao, Chong, & Liao, 2018); and monitoring of data across multi-site clinical trials (Choudhury, Fairoza, Sylla, & Das, 2019). The literature also reports various possible benefits provided by introducing blockchain in healthcare, including convenient data sharing (Zhu et al., 2019), increased data security and privacy (Liang et al., 2017; Choudhury et al., 2019), greater privacy (Khatoon, 2020), more transparency (Tseng et al., 2018), data ownership and control (Gordon & Catalini, 2018), interoperability (Wang et al., 2018), and advanced data traceability (Benchoufi, Porcher, & Ravaud, 2017).

These systems and their potential benefits could be classified in several categories, including 1) data management applications to unify and more effectively handle patient details (e.g., integration and encryption of digital assets to ensure data reliability and protection, as well as patient identity and confidentiality, patients' health records, and cloud healthcare data storage) and to facilitate data sharing between healthcare institutions, 2) supply chain management, including drugs, vaccine, and pharmaceutical supply chain, 3) internet of medical things, 4) biomedical research and education, i.e., to merge data for research purposes and minimise academic fraud, 5) health data analytics, including remote patient monitoring, supervision of drug intake, 6) health insurance claims, reporting, and fraudulent billing protection, and 7) health clinical trial analytics (Nugent, Upton, & Cimpoesu, 2016; Benchoufi et al., 2017; Ichikawa et al., 2017; Esposito et al., 2018; Agbo et al., 2019; Prokofieva & Miah, 2019).

One of the most promising applications of distributed ledger technologies in healthcare is electronic health records (EHR). An EHR is a collection of patient data "designed to allow patient medical history to move with the patient or be made available to multiple healthcare providers" (Esposito et al., 2018). EHRs allow the electronic exchange of health information with different providers to improve patients' quality of care in a meaningful way. EHRs improve clinical decision-making, reduce duplication of diagnostic testing, better medication management, increase the adoption of screening programs, and improve coordination among medical professionals (Gordon & Catalini, 2018). More broadly, other potential benefits of blockchain healthcare technology could also be seen in patient monitoring (Griggs et al., 2018). For example, Rupasinghe et al. (2019) present a conceptual BCHIS that manages and analyses EHRs to identify older adults in care who are at higher risk of falling. As another example, Uddin et al. (2018) designed an end-to-end eHealthcare framework to facilitate professional data sharing and safeguarding EHR privacy via blockchain.

The current state of aged care EHRs is disjointed due to a lack of greater interoperability between providers and hospital IS, especially concerning unified data, common architectures, and privacy and transparency concerns (Krawiec et al., 2018). The adoption of EHRs by healthcare entities is varied worldwide. Factors affecting adoption include an institution's technological infrastructure (Heart et al., 2009). Moreover, the push for personal EHRs, for which patients are more involved in health data collection (e.g., through smartphones or wearable devices) and data management, introduces further security and privacy challenges (Gordon & Catalini, 2018). As a solution to the current problems with EHRs and similar challenges with other healthcare IS, blockchain technology seeks to offer transparent and tamper-proof platform to support the integration of care-related information across a range of care and cure applications and stakeholders. The integration of blockchain technology into healthcare may improve the reliability of health data, making it resistant to tampering and revision while also enabling better accessibility and data transparency (Ichikawa et al., 2017).

Although BCHIS may bring about incredible progress in healthcare delivery, more research and co-design are needed to ensure these systems account for the diversity of values found in healthcare and avoid bias. Gender and sex considerations in healthcare are crucial because they affect individuals' health differences, yet most algorithms deployed in the healthcare context do not consider these aspects. Missing these kinds of dimensions in healthcare IS raises concern, as neglecting these aspects risks potential discrimination (Cirillo et al., 2020). Questions about the consequences of missing the gender and sex dimensions in algorithms that support decision-making processes are nevertheless particularly poorly understood and often underestimated (Buolamwini & Gebru, 2018; Fosch-Villaronga et al., 2021). Still, technology is not value-neutral and can have adverse consequences if values are not considered. This is particularly salient in healthcare because it is a sensitive domain of application with very distinct values that, if missed, can have disastrous consequences for the safety of the patients.

## 3. Method

*3.1. Value Sensitive Design*

We contribute to the literature by reflecting on the interplay between values, aged care, and distributed ledger technologies through a value sensitive design (VSD) approach. VSD is a methodology used to investigate values relating to the technological ecosystem and design systems that account for user values (Friedman & Hendry, 2019). In VSD, values refer to "what is important to people in their lives, with a focus on ethics and morality" (Friedman & Hendry, 2019). VSD seeks to empower human values in the design of technology. Today, we can see blockchain as a technology that supports human beings in a highly connected digital world as moral and prosocial persons to give and involve reliable signals of trustworthiness. In the design of new BCHIS, there is a strong need for a comprehensive theoretical and methodological framework to deal with the value dimensions of its design.

VSD considers human values during the design of new IS, or the evaluation of an old one, to resolve value conflicts and promote positive value impacts. Friedman and Hendry (2019) explain the three iterative investigations used in VSD as follows.

1. **Conceptual investigations** define the IS users and other stakeholders, identify the values of all stakeholders who interact with the IS and conceptually examine how the IS design will positively and negatively impact those values.
2. An **empirical investigation** aims to create further knowledge about those values concerning the IS through empirical means.
3. The **technical investigation** involves designing a new IS to support values as they have been understood empirically, or analysing how users interact with an existing IS.

Like other healthcare technologies, BCHIS ought to be designed for stakeholders' values, promote positive value impacts, and mitigate adverse effects. To highlight the required value and cultural sensitivity of BCHIS to avoid the replication of bias, considering the values of a specific community, in this case, the older LGBT+ community is crucial. In the following sections, we conduct a conceptual investigation of the values that LGBT+ older adults have at stake within BCHIS.

*3.2. Case study: LGBT+ older adults*

Each community has a value framework of value priorities, meanings, and orientations (Burchum, 2002), and LGBT+ older adults are no different (Waling & Roffee, 2017). Tenenbaum (2011) describes the older LGBT+ community as having distinct values, concerns, needs, and critical and experiential interests in aged care. In the search for cultural sensitivity, many LGBT+ persons seek out LGBT-friendly health services and professionals who are sensitive to their needs and values (Jann et al., 2015). The difficulty of finding a LGBT-friendly doctor leads to this group being "more likely to delay or avoid necessary medical care compared with heterosexuals" (29% versus 17%, respectively) (Khalili et al., 2015).

The older LGBT+ community's values are under-surveyed (Fredriksen-Goldsen et al., 2013), let alone their values concerning technology generally and distributed ledger healthcare technologies specifically. This lack of understanding challenges the recognition of the effects and impacts on this community, hindering these systems' potential benefits. This community's values are worthwhile investigating in BCHIS design and development to ensure they avoid discrimination and the exacerbation of existing bias against the queer community (Poulsen, Fosch-Villaronga, & Søraa, 2020; Gomes, Antonialli, & Dias-Oliva, 2019). The modicum of the literature suggests that LGBT+ older adults prioritize values of acceptance, privacy, and personhood (Tenenbaum, 2011); inclusive language and disclosing gender identity or sexual orientation (Huygen, 2006); and autonomy and empowerment (Waite, 2015). LGBT+ older adults also have different value meanings. For example, the value of family is often interpreted as a 'chosen family' consisting of close friends, rather than relatives (Cannon et al., 2017), and intersex older adults define the value of 'non-judgemental care' concerning their intersex status, as it impacts their physical, hormonal, or genetic differences (Latham & Barrett, 2015).

## 4. A Value Sensitive Design Look at Blockchain in Healthcare

Following VSD, evaluating whether a technology meets the values of a certain group requires looking at value meanings and who values them. This paper examines the use of blockchain in healthcare with the values of LGBT+ older adults in mind. The following sections draw focus to three values in particular: trust, security, and privacy. First, the emergence of these and other values in the use of blockchain in healthcare is presented. Then, the salience of those values and meanings is expanded upon. Last, focusing on the value of trust, security, and privacy, the meanings that LGBT+ older adults give to these values are explored concerning blockchain in healthcare. This last section shows how BCHIS could bring about value impacts for this community if BCHIS are not realized to account for the diversity of values found in healthcare.

*4.1. Key values in blockchain for healthcare*

In the literature, blockchain is primarily promoted as supporting trust/integrity/reliability of data (hereafter referred to collectively as trust), security, privacy, transparency, interoperability, and user control (Liang et al., 2017; Agbo et al., 2019). In addition to these being potential benefits of blockchain, several are also human values. That is, blockchain in healthcare evokes several values, including (1) patient trust in the integrity and reliability of her health data, (2) security of patient health data, and (3) privacy of patient health data.

In a survey measuring trust in health information sources, Hesse et al. (2005) found that respondents expressed a high level of trust for information provided by physicians, higher than that offered by the Internet, television, radio, family or friends, and other media. As such, the value of *trust* might relate to health information sources. Alternatively, it could relate trust in care professionals. According to the care ethics tradition, mutual trust between patients and their caregivers is essential to good care (Tronto, 2010). In care ethics, values such as trust exist within the caring relationship between patient and caregiver, not necessarily in the health information they provide, as this is implicit in the trusted relationship. Moreover, also implied in the trusted relationship is the assurance of security and privacy.

Security in healthcare can also be defined in different ways. Haas et al. (2011) draw a connection between security and access to information, suggesting that enabling "access to an increased number of users poses threats to security and privacy." In this sense, security restricts unnecessary access to information and guarantees trustworthiness. Security also relates to privacy. Without protecting health information, a patient's privacy is vulnerable (Kumar and Lee, 2012). In Europe, the General Data Protection Regulation establishes binding legal requirements to enforce the protection of these values.

In a review of the state of information security and privacy in healthcare, privacy is described as "a key governing principle of the patient-physician relationship" (Appari & Johnson, 2010). Akin to security, privacy also refers to protecting personal data, which is governed by legislation and is geared towards giving back control to users and the possibility to rectify, oppose, and cancel the processing of such data. In their proposed framework for ensuring the privacy of EHRs, Haas et al. (2011) present privacy as a driving need to securely guarantee the "controlled disclosure of personal data to third parties."

*4.2. Additional values in healthcare and technology design*

Examining blockchain in healthcare through a VSD lens, there exist other values that current blockchain technology does not make justice, although it may impact them. According to Friedman, Kahn, Borning, & Huldtgren (2013), there are some core values involved in any system design. Namely, human welfare, ownership and property, privacy, freedom from bias, universal usability, trust, autonomy, informed consent, accountability, courtesy, identity, calmness, and environmental sustainability.

The initial conceptual investigation above shows the emergence of the most salient values implicated in blockchain in healthcare – trust, security, and privacy – however, it lacks a practical examination of the diversity of values in this environment. Whereas **security** is interpreted by the technical community as an essential value, from the user's perspective and the healthcare system's goals, other values are also deemed necessary. Ensuring values are of paramount importance in healthcare. To follow is a non-comprehensive exploration of the aforementioned core values in relation to healthcare technology, including blockchain.

1. **Discrimination and bias**: Different cultures have distinct value sets, comprising different value priorities, meanings, conceptualizations, and definitions (Burchum, 2002). Culture shapes a person's value system; one of the most significant influences on an individual's values is each person's cultural background (Liu et al., 2014; Sunny et al., 2019). Cultural knowledge, i.e., understanding cultural values, is key to cultural competence in care, including technology-driven care (Tsaur & Tu, 2019).
2. **Consent and autonomy**: Although distributed ledger technologies provide a "shared, immutable, and transparent audit trail for accessing data" (Rupasinghe et al., 2019), this data is used on many occasions as a basis for predictive analysis, which has several associated ethical, legal, and societal concerns (Amarasingham et al., 2014). Predictive analytics refers to the use of statistical models to determine future performance based on current and historical data and raise questions in the context of healthcare. These questions range from how patients gave meaningful consent. What if harm could have been averted if physicians would study their patients more carefully instead of relying on predictive models (Hoffman, & Podgurski, 2009)? The legal research is also rich in pointing out the risks to privacy and data protection in this regard, even in connection to the fairness, accountability, and transparency of the models, for inferences and decisions made upon predictive algorithmic decision-making (Wachter & Mittelstadt, 2019). Making information open and transparent requires the affected individuals to understand and assess the risks of predictive analytics and automated decision-making systems, allowing them to challenge decisions concerning them (Felzmann et al., 2019).
3. The fact that distributed ledgers eradicate the middleman also blur the **accountability** linked to them. The implementation of blockchain must ensure that a group of persons, institutions, or legal entities are accountable for their actions and assume responsibility.
4. **Courtesy:** In healthcare, good care practice requires cultural competence, value sensitivity, and person-centredness (Purnell & Fenkl, 2019; Santana et al., 2018). Since technology is not value-neutral (Friedman & Hendry, 2019), healthcare technologies also need to demonstrate value sensitivity and cultural competence. Good cultural competence in the provision of care is value-based (Markey & Okantey, 2019), and thus cultural competence in healthcare technologies could be achieved with a value sensitive design approach (Poulsen & Burmeister, 2019, Poulsen et al., 2018).

5. **Identity:** The value of identity is complicated. One's identity is respected in that the person retains personal data ownership using blockchain. However, at the same time, parties are pseudonymous, so one's identity is partially lost on the blockchain.

## 5. LGBT+ Older Adult Values and Blockchain in Healthcare: Overlooked Interpretations of Trust, Security, and Privacy

In this section, we reflect on how blockchain in healthcare conceptually impacts the values of the older LGBT+ community. Here, we draw focus on the values of trust, security, and privacy.

In an LGBT+ aged care study, Willis et al. (2016) identified several critical interpretations of privacy, finding that this community values private time with partners and friends, having partners and friends feel welcome, and acknowledging younger-older LGBT+ partnerships. These interpretations of privacy are particularly important given the prevalence of chosen families and the lack of education among health professionals on the importance of nonrelatives as a source of support for the older LGBT+ community (Cannon et al., 2017), which could result in breaches of privacy with their partners and friends in a secure, safe space. The lack of education on these topics raises the additional concern of trust in health professionals; without LGBT-friendly service provision, a level of trust is lost.

LGBT+ older adults may interpret the value of security as that supported by safe spaces. These environments enable LGBT+ older adults to safely disclose their status with like-minded people (Hughes, Harold, & Boyer, 2011). Healthcare environments that display symbols of LGBT+ acceptance, such as rainbow flags and pictures of same-sex couples, help provide a sense of feeling safe (Willis et al., 2016). More on the value of security for the older LGBT+ community, Crameri et al. (2015) highlights the importance of being aware of the histories of older LGBT+ people and being educated in this area to provide security for those in care.

Reflecting on the older LGBT+ community's interpretation of security as safe spaces, examine the following potential negative value impact if this conceptualization is not understood in the realization of BCHIS. Consider a gay man living with dementia who has challenges remembering he is in a same-sex partnership. When this person is made aware that he has a partner, he gets confused and feels uncomfortable being reminded of persisting sexual identity issues emerging from historical discrimination. Now consider this person's partner who takes them to doctor visits. They prefer that health professionals do not greet them as partners because it will make their partner living with dementia upset. The value of security for these LGBT+ older adults is interpreted as the safe space that their regular doctor's office creates in this way. Concerning the effective interoperability of blockchain data, if this same-sex couple visits a new healthcare facility made aware that a gay couple is visiting due to the shared health information and, thus, they greet them as such, their value of security is negatively impacted.

On the other hand, consider the positive value impact of accounting the security as conceptualized as 'safely disclosing LGBT+ status and feeling safe.' Regarding the blockchain-enabled management of EHRs, LGBT+ older adults need only disclose their gender or sexuality once, if it is relevant to their health assessment. EHRs will be retained and not misplaced on the decentralized blockchain, satisfying the value of security. Whereas health records kept on paper can be lost and an LGBT+ older adult may need to disclose their status again, potentially making them feel unsafe.

Reflecting on historical discrimination and the importance of ensuring that LGBT+ status remains private for those who value it (Crameri et al., 2015), consider the negative value impact of failing to interpret privacy as 'private LGBT+ status' in BCHIS design. Due to historical discrimination, the LGBT+ community highly values privacy related to gender or sexuality for fear of social stigma. Health information stored with blockchain is irreversible without consensus. This irreversibility creates a problem for those LGBT+ older adults who do not want their gender or sexuality to be shared with care services. Moreover, this problem might be exacerbated if an LGBT+ older adult changes their sexuality or gender as they age and want to erase previous identifiers on EHRs.

A systematic review of the perceptions of older LGBT+ people regarding sexuality in residential healthcare by Mahieu, Cavolo, and Gastmans (2019) identified that a lack of privacy is of significant concern. Mahieu et al. (2019) raise the issue that a perceived lack of privacy in residential care facilities would prevent socialization and sexual expression. Similar problems with trust are raised here. Without appropriate LGBT-friendly measures being put in place by health services, a lack of trust arises. Villar et al. (2014) similarly affirm the value of private time with partners and friends and safe spaces to enable a level of security for LGBT+ older adults, stating that "measures designed to increase residents' opportunities for privacy and intimacy, in terms of both time and space, seem necessary." The authors highlight that private spaces, less control, trust, personal LGBT+ relationships, and safe spaces are essential values for the older LGBT+ community. Consider the potential positive value impact of ensuring that the value of privacy is conceptualized as 'private LGBT+ relationships' in creating BCHIS. For some LGBT+ older adults, intimate relationships are valuable. If BCHIS provide a more reliable private communication network than existing IS, then LGBT+ older adults wishing to hide their relationship out of fear of stigma might be put at ease by higher levels of security and privacy of health information.

Much of the literature about what privacy and trust mean to the older LGBT+ community concerns relationships, intimacy, and sexual expression. Yet, there are other needs relating to trust and privacy that are distinct for LGBT+ older adults. Ansara (2015) notes that service providers ought to inspire trust in service provision by creating "a privacy policy for people who have previously received services in another gender." That is, the value of preferred gender affirmation is an essential interpretation for both privacy and trust in this community. Reflecting on a potentially negative value impact, consider the failure to understand that the older LGBT+ community interprets trust as 'confidence in LGBT-friendly physicians and preferred gender affirmation.' Trust in an LGBT-friendly physician is essential for LGBT+ older adults to ensure that one's gender or sexuality is respected when a health assessment is made. For instance, an older transman might interpret trust as that which exists in the private, unspoken agreement with their doctor that they will acknowledge their biological gender in their health assessments but will always address them by their preferred gender. This trust extends to the assurance that the doctor will not share this information with others, including other health professionals or care providers. Using blockchain to efficiently share health information might negatively impact this LGBT+ older adult's value of trust as all networked care services will know their transgender status even though they might not need to know.

Furthermore, LGBT+ persons value a level of trust in LGBT-friendly healthcare professionals who can accurately report on LGBT-related health information (Jann et al., 2015; Khalili et al., 2015), such as one's preferred gender. Consider the positive value impacts for LGBT+ older adults by ensuring that trust is effectively interpreted as 'confidence in the accuracy of LGBT-related health information' in BCHIS design. Consider an LGBT+ older adult with a history of having doctors who fail to account for their distinct LGBT+ needs when making medical assessments. If blockchain does provide a platform for maintaining more accurate health information than alternate IS, then this person's value of trust will be promoted.

Current blockchain technology does not consider these values and value meanings. Presented often as value-neutral, blockchain is usually praised for enabling a more transparent, open, and power-distributed internet. Still, it has the potential of replicating existing biases against specific communities.

## 6. Discussion

The purpose of this study is to address the implications of the use and development of blockchain, for specific vulnerable populations, such as older adults, persons with disabilities, or children, in healthcare settings without integrating human values, ethical considerations, or legal aspects. Here, we focus on the older LGBT+ community to emphasize the need to account for the diversity of values implicated in the design of BCHIS, which include different sets of value meanings. Accounting for diversity may avoid replicating biases into systems, increasing trust in the system, and ensuring that other values connected to privacy and security are ensured. Applying value sensitive design (VSD), this paper shows the different values that need to be taken into account for the deployment of ledger technologies for healthcare, as these can have positive and negative impacts on the values of aging populations. Mainly, trust, security, and privacy layers are crucial at the personal and community levels.

Accounting for the diversity of values found in care should translate into technology, including distributed ledger technologies. However, this change, although particularly salient in sensitive sectors such as healthcare, the introduction of blockchain in this space requires a "radical rethink and significant investment in the entire ecosystem" (Esposito et al., 2018). Moreover, from the same blockchain, new values arise, and further impacts will need to be identified, explored, and resolved (Pazaitis, De Filippi, & Kostakis, 2017). Resolved, in this sense, means that positive impacts should be encouraged, whereas negative impacts should be reconciled.

Discrimination and bias are inherent problems of many systems, including blockchain, AI, and robots (Raji & Buolamwini, 2019). Many biases in the offline world may propagate to systems that heavily rely on human input if not addressed. For example, gender classification systems, often used by social media platforms to infer the gender of users for advertising and personalization purposes, are trained on real-world datasets and are often biased because the data used to train them is biased, containing namely racial and gender stereotypes (Torralba & Efros, 2011; Buolamwini, & Gebru, 2018; Font & Costa-jussa, 2019; McDuff et al., 2019). Zhao et al. (2017) found that the datasets imSitu and MS-COCO used to train gender classification systems are significantly gender-biased and that "models trained to perform prediction on these datasets amplify the existing gender bias when evaluated on development data." For example, the verb 'cooking' is heavily biased towards females in a system trained using the imSitu dataset, amplifying existing gender biases (Zhao et al., 2017). The same gender biases have been shown in natural language processing (Zhou et al., 2019), another method used to support gender classification systems (Campa, Davis, & Gonzalez, 2019).

Another example of system bias replication is "statistical discrimination," which refers to making (un)educated guesses about an unobservable candidate characteristic, such as which applicants perform well as employees. This has been proven quite problematic from the Amazon-hiring algorithm failure, where women candidates were more often devalued than men, as the company traditionally had hired few women (Bogen, 2019). The algorithm concluded that being a woman was an undesirable characteristic for recruitment purposes. Thus, having a curriculum vitae with the entry of being president of the "women's chess club" was seen as a red flag, giving the candidate more negative scores, while just generally being a member of a "chess club" was seen as positive.

Understanding the impact that algorithms have on different communities and values is challenging, as they may appear much later, usually after being widely used (Hao, 2019), but undoubtedly necessary to make a fairer society. It is often the case that those communities mostly remain "invisible, silent, powerless, and unable to understand how these technologies may affect them" (Poulsen, Fosch-Villaronga, & Søraa, 2020).

## 7. Conclusions

Our contribution shows that while blockchain developers usually consider the values trust, security, and privacy when designing blockchain, they do not usually acknowledge and integrate other human values such as human welfare, ownership and property, privacy, freedom from bias, universal usability, trust, autonomy, informed consent, accountability, courtesy, identity, calmness, and environmental sustainability. In this paper, we also stress that these are not *mere* values to be integrated into the design of a system. Values need to be juxtaposed to the meaning specific communities ascribe to those values. In the paper, we refer to the LGBT+ literature to reflect on how LGBT+ older adults interpret these values distinctly.

By highlighting under-surveyed values of LGBT+ older adults, our contribution alerts us to the potential discriminatory implications of these technologies, which do not consider vulnerable, silent populations. As such, distributed ledger technology developers should reflect on the effects their work on efficiency, privacy, and security, and user values, and of the importance of acting upon and integrating those reflections timely and adequately in the development of a new technology aligns with *Responsible Innovation* (Stilgoe, Owen, & Macnaghten, 2013). Focusing on one community throughout, the older LGBT+ community, we emphasize the need for a holistic, value sensitive design approach for the development of ledger technologies for healthcare, which include the values of everyone within the healthcare ecosystem.

**Acknowledgements**

EFV would like to acknowledge that contributions to this research are supported by the European Research Council Starting Grant SAFE and SOUND project, which received funding from the European Union's Horizon-ERC program Grant Agreement No. 101076929. Views and opinions expressed are however those of the authors only and do not necessarily reflect those of the European Union or the European Research Council. Neither the European Union nor the granting authority can be held responsible for them.

**References**

Agbo, C. C., Mahmoud, Q. H., & Eklund, J. M. (2019). Blockchain technology in healthcare: A systematic review. *Healthcare, 7*(2), 1-30. https://doi.org/10.3390/healthcare7020056

Amarasingham, R., Patzer, R. E., Huesch, M., Nguyen, N. Q., & Xie, B. (2014). Implementing electronic health care predictive analytics: Considerations and challenges. *Health Affairs, 33*(7), 1148-1154. https://doi.org/10.1377/hlthaff.2014.0352

Andoni, M., Robu, V., Flynn, D., Abram, S., Geach, D., Jenkins, D., ... & Peacock, A. (2019). Blockchain technology in the energy sector: A systematic review of challenges and opportunities. *Renewable and Sustainable Energy Reviews, 100*, 143-174. https://doi.org/10.1016/j.rser.2018.10.014

Appari, A., & Johnson, M. E. (2010). Information security and privacy in healthcare: Current state of research. *Int. J. Internet and Enterprise Management, 6*(4), 279-314. https://doi.org/10.1504/IJIEM.2010.035624

Atzei, N., Bartoletti, M., & Cimoli, T. (2017). A survey of attacks on Ethereum smart contracts (SoK). In M. Maffei, & M. Ryan (Eds.), *Principles of security and trust. POST2017* (pp. 164-186). Springer. https://doi.org/10.1007/978-3-662-54455-6_8

Raji, I. D. & Buolamwini, J. (2019). Actionable auditing: Investigating the impact of publicly naming biased performance results of commercial AI products. In *Proceedings of the Conference on Artificial Intelligence, Ethics, and Society* (pp. 429-435). AAAI/ACM. https://www.media.mit.edu/publications/actionable-auditing-investigating-the-impact-of-publicly-naming-biased-performance-results-of-commercial-ai-products/

Benchoufi, M., Porcher, R., & Ravaud, P. (2017). Blockchain protocols in clinical trials: Transparency and traceability of consent. *F1000Research, 6*, 1-72. https://doi.org/10.12688/f1000research.10531.5

Bogen, M. (2019). *All the ways hiring algorithms can introduce bias*. Harvard Business Review. https://hbr.org/2019/05/all-the-ways-hiring-algorithms-can-introduce-bias

Buolamwini, J., & Gebru, T. (2018). Gender shades: Intersectional accuracy disparities in commercial gender classification. *Proceedings of Machine Learning Research, 81*, 77-91. https://proceedings.mlr.press/v81/buolamwini18a.html

Burchum, J. L. R. (2002). Cultural competence: An evolutionary perspective. *Nursing Forum, 37*(4), 5-15. https://doi.org/10.1111/j.1744-6198.2002.tb01287.x

Campa, S., Davis, M., & Gonzalez, D. (2019). Deep & machine learning approaches to analyzing gender representations in journalism. https://web.stanford.edu/class/archive/cs/cs224n/cs224n.1194/reports/custom/15787612.pdf

Cannon, S. M., Shukla, V., & Vanderbilt, A. A. (2017). Addressing the healthcare needs of older lesbian, gay, bisexual, and transgender patients in medical school curricula: A call to action. *Med Educ Online, 22*(1), 1-4. https://doi.org/10.1080%2F10872981.2017.1320933

Choudhury, O., Fairoza, N., Sylla, I., & Das, A. (2019). *A blockchain framework for managing and monitoring data in multi-site clinical trials*. ArXiv. https://doi.org/10.48550/arXiv.1902.03975
18

Cirillo, D., Catuara-Solarz, S., Morey, C. et al. (2020). Sex and gender differences and biases in artificial intelligence for biomedicine and healthcare. *NPJ Digital Medicine, 3*, 1-11. https://doi.org/10.1038/s41746-020-0288-5

Cresswell, K., Cunningham-Burley, S., & Sheikh, A. (2018). Health care robotics: Qualitative exploration of key challenges and future directions. *Journal of Medical Internet Research, 20*(7), 1-11. https://doi.org/10.2196/10410

De Filippi, P., & Loveluck, B. (2016). The invisible politics of Bitcoin: Governance crisis of a decentralized infrastructure. *Internet Policy Review, 5*(3), 1-28. https://doi.org/10.14763/2016.3.427

Esposito, C., Santis, A. D., Tortora, G., Chang, H., & Choo, K. R. (2018). Blockchain: A panacea for healthcare cloud-based data security and privacy? *IEEE Cloud Computing, 5*(1), 31-37. https://doi.org/10.1109/MCC.2018.011791712

Felzmann, H., Fosch-Villaronga, E., Lutz, C., & Tamò-Larrieux, A. (2019). Transparency you can trust: Transparency requirements for artificial intelligence between legal norms and contextual concerns. *Big Data & Society*, *6*(1), 1-14. https://doi.org/10.1177/2053951719860542

Font, J., & Costa-jussa, M. (2019). Equalizing gender bias in neural machine translation with word embeddings techniques. In *Proceedings of the First Workshop on Gender Bias in Natural Language Processing* (pp. 147-154). ACL. http://dx.doi.org/10.18653/v1/W19-3821

Fosch-Villaronga, E., Poulsen, A., Søraa, R. A., & Custers, B. H. M. (2021). A little bird told me your gender: Gender inferences in social media. *Information Processing & Management*, *58*(3), 1-13. https://doi.org/10.1016/j.ipm.2021.102541

Friedman, B., & Hendry, D. G. (2019). Value sensitive design: Shaping technology with moral imagination. MIT Press. https://doi.org/10.7551/mitpress/7585.001.0001

Friedman, B., Kahn, P. H., Borning, A., & Huldtgren, A. (2013). Value sensitive design and information systems. In N. Doorn, D. Schuurbiers, I. van de Poel, & M. E. Gorman (Eds.), *Early engagement and new technologies: Opening up the laboratory* (pp. 55-95). Springer. https://doi.org/10.1007/978-94-007-7844-3_4

Frizzo-Barker, J., Chow-White, P. A., Adams, P. R., Mentanko, J., Ha, D., & Green, S. (2020). Blockchain as a disruptive technology for business: A systematic review. *International Journal of Information Management, 51*, 1-14. https://doi.org/10.1016/j.ijinfomgt.2019.10.014

Gomes, A., Antonialli, D. & Dias-Oliva, T. (2019). *Drag queens and artificial intelligence: Should computers decide what is toxic on the internet?* Internet Lab blog. http://www.internetlab.org.br/en/freedom-of-expression/drag-queens-and-artificial-intelligence-should-computers-decide-what-is-toxic-on-the-internet/

Gordon, W. J., & Catalini, C. (2018). Blockchain technology for healthcare: Facilitating the transition to patient-driven interoperability. *Computational and Structural Biotechnology Journal, 16*, 224-230. https://doi.org/10.1016/j.csbj.2018.06.003

Griggs, K. N., Ossipova, O., Kohlios, C. P., Baccarini, A. N., Howson, E. A., & Hayajneh, T. (2018). Healthcare blockchain system using smart contracts for secure automated remote patient monitoring. *J Med Syst, 42*(7), 1-7. https://doi.org/10.1007/s10916-018-0982-x

Haas, S., Wohlgemuth, S., Echizen, I., Sonehara, N., & Müller, G. (2011). Aspects of privacy for electronic health records. *International Journal of Medical Informatics, 80*(2), e26-e31. https://doi.org/10.1016/j.ijmedinf.2010.10.001

Hao, K. (2019). *This is how AI bias really happens—and why it's so hard to fix*. MIT Technology Review. https://www.technologyreview.com/2019/02/04/137602/this-is-how-ai-bias-really-happensand-why-its-so-hard-to-fix/

Heart, T., O'Reilly, P., Sammon, D., & O'Donoghue, J. (2009). Bottom-up or top-down? *Journal of Systems and Information Technology, 11*(3), 244-268. https://doi.org/10.1108/13287260910983623

Hesse, B. W., Nelson, D. E., Kreps, G. L., Croyle, R. T., Arora, N. K., Rimer, B. K., & Viswanath, K. (2005). Trust and sources of health information: The impact of the internet and its implications for health care providers: Findings from the first health information national trends survey. *JAMA Internal Medicine, 165*(22), 2618-2624. https://doi.org/10.1001/archinte.165.22.2618

Hoffman, S., & Podgurski, A. (2009). E-health hazards: Provider liability and electronic health record systems. *Berkeley Technology Law Journal*, *24*(4), 1523-1581. https://www.jstor.org/stable/24120587

Hoy, M. B. (2017). An introduction to the blockchain and its implications for libraries and medicine. *Medical Reference Services Quarterly*, *36*(3), 273-279. https://doi.org/10.1080/02763869.2017.1332261

Hughes, M. (2016). Loneliness and social support among lesbian, gay, bisexual, transgender and intersex people aged 50 and over. *Ageing and Society, 36*(9), 1961-1981. https://doi.org/10.1017/S0144686X1500080X

Hughes, L., Dwivedi, Y. K., Misra, S. K., Rana, N. P., Raghavan, V., & Akella, V. (2019). Blockchain research, practice and policy: Applications, benefits, limitations, emerging research themes and research agenda. *International Journal of Information Management, 49*, 114-129. https://doi.org/10.1016/j.ijinfomgt.2019.02.005

Huygen, C. (2006). Understanding the needs of lesbian, gay, bisexual, and transgender people living with mental illness. *Medscape General Medicine, 8*(2), 1-6. http://www.ncbi.nlm.nih.gov/pmc/articles/pmc1785208/

Iansiti, M., & Lakhani, K. R. (2017). *The truth about blockchain*. Harvard Business Review. https://hbr.org/2017/01/the-truth-about-blockchain

Ichikawa, D., Kashiyama, M., & Ueno, T. (2017). Tamper-resistant mobile health using blockchain technology. *JMIR Mhealth Uhealth, 5*(7), 1-11. https://doi.org/10.2196/mhealth.7938

Jann, J. T., Edmiston, E. K., & Ehrenfeld, J. M. (2015). Important considerations for addressing LGBT health care competency. *American Journal of Public Health, 105*(11). https://doi.org/10.2105%2FAJPH.2015.302864

Khalili, J., Leung, L. B., & Diamant, A. L. (2015). Finding the perfect doctor: Identifying lesbian, gay, bisexual, and transgender-competent physicians. *Am J Public Health, 105*(6), 1114-1119. https://doi.org/10.2105/AJPH.2014.302448

Khatoon, A. (2020). A blockchain-based smart contract system for healthcare management. *Electronics, 9*(1), 1-23. https://doi.org/10.3390/electronics9010094

Krawiec, R. J., Housman, D., White, M., Filipova, M., Quarre, F., Barr, D., . . . L., T. (2018). *Blockchain: Opportunities for health care*. Deloitte. https://www2.deloitte.com/us/en/pages/public-sector/articles/blockchain-opportunities-for-health-care.html

Kumar, P., & Lee, H.-J. (2012). Security issues in healthcare applications using wireless medical sensor networks: A survey. *Sensors, 12*(1), 55-91. https://doi.org/10.3390/s120100055

Kuo, T.-T., Kim, H.-E., & Ohno-Machado, L. (2017). Blockchain distributed ledger technologies for biomedical and health care applications. *Journal of the American Medical Informatics Association, 24*(6), 1211-1220. https://doi.org/10.1093/jamia/ocx068

Latham, J., & Barrett, C. (2015). Appropriate bodies and other damn lies: Intersex ageing and aged care. *Australasian Journal on Ageing, 34*(S2), 19-20. https://doi.org/10.1111/ajag.12275

Liang, X., Zhao, J., Shetty, S., Liu, J., & Li, D. (2017). Integrating blockchain for data sharing and collaboration in mobile healthcare applications. In *Proceedings of the 2017 IEEE 28th Annual International Symposium on Personal, Indoor, and Mobile Radio Communications (PIMRC)* (pp. 1-5). IEEE. https://doi.org/10.1109/PIMRC.2017.8292361

Liu, S., Volcic, Z., & Gallois, C. (2014). *Introducing intercultural communication: Global cultures and contexts* (4th ed.). SAGE.

Markey, K., & Okantey, C. (2019). Nurturing cultural competence in nurse education through a values-based learning approach. *Nurse Education in Practice, 38*, 153-156. https://doi.org/10.1016/j.nepr.2019.06.011

McDuff, D., Song, Y., Kapoor, A., & Ma, S. (2019). Characterizing bias in classifiers using generative models. In *Proceedings of the 33rd International Conference on Neural Information Processing Systems* (pp. 5403-5414). ACM. https://dl.acm.org/doi/abs/10.5555/3454287.3454772

Nugent, T., Upton, D., & Cimpoesu, M. (2016). Improving data transparency in clinical trials using blockchain smart contracts. *F1000Research*, *5*, 1-9. https://doi.org/10.12688/f1000research.9756.1

Pazaitis, A., De Filippi, P., & Kostakis, V. (2017). Blockchain and value systems in the sharing economy: The illustrative case of Backfeed. *Technological Forecasting and Social Change*, *125*, 105-115. https://doi.org/10.1016/j.techfore.2017.05.025

Poulsen, A., & Burmeister, O. K. (2019). Overcoming carer shortages with care robots: Dynamic value trade-offs in run-time. *Australasian Journal of Information Systems, 23*, 1-18. https://doi.org/10.3127/ajis.v23i0.1688

Poulsen, A., Burmeister, O. K., & Tien, D. (2018). A new design approach and framework for elderly care robots. In *Proceedings of the Australasian Conference on Information Systems* (pp. 1-12). UTS ePRESS. https://doi.org/10.5130/acis2018.cj

Poulsen, A., Fosch-Villaronga, E., & Søraa, R. A. (2020). Queering machines. *Nature Machine Intelligence, 2*. https://doi.org/10.1038/s42256-020-0157-6

Prokofieva, M., & Miah, S. J. (2019). Blockchain in healthcare. *Australasian Journal of Information Systems, 23*, 1-22. https://doi.org/10.3127/ajis.v23i0.2203

Purnell, L. D., & Fenkl, E. A. (2019). The Purnell Model for cultural competence. In *Handbook for culturally competent care* (pp. 7-18). Springer. https://doi.org/10.1007/978-3-030-21946-8_2

Queiroz, M. M., & Wamba, S. F. (2019). Blockchain adoption challenges in supply chain: An empirical investigation of the main drivers in India and the USA. *International Journal of Information Management, 46*, 70-82. https://doi.org/10.1016/j.ijinfomgt.2018.11.021

Rupasinghe, T., Burstein, F., Rudolph, C., & Strange, S. (2019). Towards a Blockchain based Fall Prediction Model for Aged Care. In *Proceedings of the Australasian Computer Science Week Multiconference* (pp. 1-10). ACM. https://doi.org/10.1145/3290688.3290736

Santana, M. J., Manalili, K., Jolley, R. J., Zelinsky, S., Quan, H., & Lu, M. (2018). How to practice person-centred care: A conceptual framework. *Health Expectations, 21*(2), 429-440. https://doi.org/10.1111/hex.12640

Schuetz, S., & Venkatesh, V. (2020). Blockchain, adoption, and financial inclusion in India: Research opportunities. *International Journal of Information Management*, *52*, 1-8. https://doi.org/10.1016/j.ijinfomgt.2019.04.009

Stilgoe, J., Owen, R., & Macnaghten, P. (2013). Developing a framework for responsible innovation. *Research Policy,42*(9), 1568-1580. https://doi.org/10.1016/j.respol.2013.05.008

Sunny, S., Patrick, L., & Rob, L. (2019). Impact of cultural values on technology acceptance and technology readiness. *International Journal of Hospitality Management, 77*, 89-96. https://doi.org/10.1016/j.ijhm.2018.06.017

Tenenbaum, E. M. (2011). Sexual expression and intimacy between nursing home residents with dementia: Balancing the current interests and prior values of heterosexual and LGBT residents. *Temple Political and Civil Rights Law Review, 21*, 459-483. https://ssrn.com/abstract=2149841

Torralba, A., & Efros, A. A. (2011). Unbiased look at dataset bias. In *Proceedings of the Computer Vision and Pattern Recognition (CVPR)* (pp. 1521-1528). IEEE. https://doi.org/10.1109/CVPR.2011.5995347

Tronto, J. C. (2010). Creating caring institutions: Politics, plurality, and purpose. *Ethics & Social Welfare, 4*(2), 158-171. https://doi.org/10.1080/17496535.2010.484259

Tsaur, S.-H., & Tu, J.-H. (2019). Cultural competence for tour leaders: Scale development and validation. *Tourism Management, 71*, 9-17. https://doi.org/10.1016/j.tourman.2018.09.017

Tseng, J. H., Liao, Y. C., Chong, B., & Liao, S. W. (2018). Governance on the Drug Supply Chain via Gcoin Blockchain. *Int J Environ Res Public Health, 15*(6), 1-8. https://doi.org/10.3390%2Fijerph15061055

Uddin, M. A., Stranieri, A., Gondal, I., & Balasubramanian, V. (2018). Continuous patient monitoring with a patient centric agent: A block architecture. *IEEE Access, 6*, 32700-32726. https://doi.org/10.1109/ACCESS.2018.2846779

Wachter, S., & Mittelstadt, B. (2019). A right to reasonable inferences: Re-thinking data protection law in the age of big data and AI. *Columbia Business Law Review, 2019*(2), 1-130. https://ssrn.com/abstract=3248829

Waite, H. (2015). Old lesbians: Gendered histories and persistent challenges. *Australasian Journal on Ageing, 34*(S2), 8-13. https://doi.org/10.1111/ajag.12272

Waling, A., & Roffee, J. A. (2017). Knowing, performing and holding queerness: LGBT+ student experiences in Australian tertiary education. *Sex Education, 17*(3), 302-318. https://doi.org/10.1080/14681811.2017.1294535

Wang, S., Wang, J., Wang, X., Qiu, T., Yuan, Y., Ouyang, L., . . . Wang, F. (2018). Blockchain-powered parallel healthcare systems based on the ACP approach. *IEEE Transactions on Computational Social Systems, 5*(4), 942-950. https://doi.org/10.1109/TCSS.2018.2865526

Warkentin, M., & Orgeron, C. (2020). Using the security triad to assess blockchain technology in public sector applications. *International Journal of Information Management, 52,* 1-8. https://doi.org/10.1016/j.ijinfomgt.2020.102090

Wright, A., & De Filippi, P. (2015). Decentralized blockchain technology and the rise of Lex Cryptographia. *Social Science Research Network*, 1-58. https://dx.doi.org/10.2139/ssrn.2580664

Zhang, P., Schmidt, D. C., White, J., & Lenz, G. (2018). Blockchain technology use cases in healthcare. *Advances in Computers, 111*, 1-41. https://doi.org/10.1016/bs.adcom.2018.03.006

Zhao, J., Wang, T., Yatskar, M., Ordonez, V., & Chang, K-W. (2017). Men also like shopping: Reducing gender bias amplification using corpus-level constraints. In *Proceedings of the 2017 Conference on Empirical Methods in Natural Language Processing* (pp. 2979-2989). ACL. http://dx.doi.org/10.18653/v1/D17-1323

Zhou, P., Shi, W., Zhao, J., Huang, K-H., Chen, M., Cotterell, R., & Chang, K-W. (2019). Examining gender bias in languages with grammatical gender. In *Proceedings of the 2019 Conference on Empirical Methods in Natural Language Processing and the 9th International Joint Conference on Natural Language Processing (EMNLP-IJCNLP)* (pp. 5279-5287). ACL. http://dx.doi.org/10.18653/v1/D19-1531

Zhu, X., Shi, J., & Lu, C. (2019). Cloud health resource sharing based on consensus-oriented blockchain technology: Case study on a breast tumor diagnosis service. *J Med Internet Res, 21*(7), 1-15. https://doi.org/10.2196%2F13767